\documentclass[12pt]{article}

\usepackage{amsmath,amssymb,amsthm,latexsym} 
\usepackage{amsfonts}
\usepackage[english]{babel} 
\usepackage{graphicx,color} 

\usepackage{array}
\usepackage{mathtools}

\newtheorem{lemma}{Lemma}[section]

\newcommand{\bbone}{{\bf{1}}}

\newtheorem{theorem}{Theorem}[section]
\newcommand{\bea}{\begin{eqnarray}}
\newcommand{\eea}{\end{eqnarray}}
\newcommand{\bee}{\begin{equation}}
\newcommand{\ee}{\end{equation}}

\newcommand{\Tr}{{\rm Tr}}

\newcommand{\cH}{{\cal{H}}}

\newcommand{\gF}{{\mathfrak{F}}}
\newcommand{\gT}{{\mathfrak{T}}}

\newcommand{\cS}{{\cal{S}}}
\newcommand{\LV}{\cS}
\newcommand{\cT}{{\cal{T}}}

\newcommand{\cF}{{\cal{F}}}

\newcommand{\prf}{{\noindent {\rm \bf Proof}\, }}

\def\spvecA#1;{\if;#1;\else #1\cr \expandafter \spvecA \fi}

\numberwithin{equation}{section}

\begin{document} 

\title{Constructive Matrix Theory\\ for Higher Order Interaction\\
II: Hermitian and Real Symmetric Cases}
\author{
Thomas Krajewski$^{\star}$, Vincent Rivasseau$^{\sharp}$, Vasily Sazonov$^{\flat,\sharp}$\\
\\
$^{\star}\; $Centre de Physique Th\'eorique, CNRS UMR 7332\\
Universit\'e Aix-Marseille,
F-13009 Marseille  \\
$^{\sharp}$ Laboratoire de Physique Th\'eorique, CNRS UMR 8627,\\ 
Universit\'e Paris-Sud,  F-91405 Orsay,\\
$^{\flat}$ Institute of Physics, University of Graz,\\ Universit\"atsplatz 5, 8010, Graz, Austria}

\maketitle 

\begin{abstract} 
This paper provides the constructive \emph{loop vertex expansion}
for stable matrix models with (single trace) interactions of arbitrarily high even order in the Hermitian and real symmetric cases. 
It relies on a new and simpler method which can also be applied in the previously treated complex case. 
We prove analyticity in the coupling constant of the free energy for such models
in a domain \emph{uniform in the size $N$ of the matrix}. 
\end{abstract} 

\begin{flushright}
LPT-20XX-xx
\end{flushright}
\medskip


\vfill\eject

\section{Introduction}

In this sequel to \cite{Krajewski:2017thd} we extend the analyticity results  on $N$  by $N$ 
complex matrix models
to the case of \emph{Hermitian} or \emph{real symmetric} matrices with a higher than quartic positive even interaction\footnote{Presumably our method extends also to the ten Atland-Zirnbauer
discrete symmetry classes \cite{AZ}.}. 
Such models are interesting for many areas of theoretical physics, in particular in the context of two dimensional quantum gravity. Notice that Feynman graphs of Hermitian matrix models pave orientable Riemann surfaces of arbitrary genus and in the real symmetric case also pave \emph{non-orientable surfaces}.

Since this paper is a sequel to \cite{Krajewski:2017thd} we refer to the latter introduction for further explanation on our program and motivation. But we would like to stress that the improved method introduced in this paper is both \emph{simpler and more powerful}. The basic formalism is still the Loop Vertex Representation
or LVR\footnote{This LVR representation is itself a
generalization of the Loop Vertex Expansion \cite{Rivasseau:2007fr}. The latter is well-adapted only to quartic interactions (see \cite{Erbin:2019zug} for a recent review).} first introduced in \cite{Rivasseau:2017hpg}, joined to Cauchy holomorphic matrix calculus as in \cite{Krajewski:2017thd}. But when \cite{Krajewski:2017thd} used
contour integral parameters attached to every \emph{vertex} of the loop representation, this paper introduces more contour integrals, one for each loop vertex \emph{corner}.
This results in simpler bounds for the norm of the corner operators. 

In the scalar case $N=1$ \cite{Rivasseau:2007fr} the corresponding 
analytic contour integrals and bounds 
reduce to some ``poor man" particular case of the general theory of resurgent 
calculus of Jean \'Ecalle and followers \cite{ecalle}-\cite{ecallemenous}. See in this respect \cite{Fauvetetal} for a recent reference
on scalar partition functions in zero dimension. However the emphasis in this paper as in \cite{Krajewski:2017thd}
is on obtaining \emph{uniform} bounds for the matrix free energies as $N \to \infty$. 

Let us in this respect also emphasize that the LVE expresses the partition function of a quantum field theory as a sum over a weighted combinatorial species of decorated \emph{forests}, which has the same advantage than the traditional expansion in terms of Feynman graphs, namely its logarithm or free energy is given by the \emph{same sum} but restricted to the corresponding \emph{connected species of trees},
but has in addition the great advantage of being a \emph{convergent sum}.

\section{Hermitian Case}
Let  $d \mu (H) $ be the standard normalized GUE measure with 
iid covariance $1/N$ between matrix elements
so that 
\bee  d\mu (H) = \frac{1}{\pi^{N^2}} e^{-N\Tr H^2} dH ,
\ee
where $dH = \prod_i dH_{ii} \prod_{i<j}  dH_{ij}d\bar H_{ij}$.
We consider the 
Hermitian matrix model with stable interaction of order $2p$ with $p \ge 2$
\bee S_0(\lambda, H) := \lambda \Tr H^{2p} ,
\ee
where $\lambda$ is the coupling constant. Remark that the case
$p=2$ is much simpler than the general case $p\ge 3$, and  has been first treated with the help of the intermediate field representation in \cite{Rivasseau:2007fr}.
The partition function and free energy of the model are given by
\begin{eqnarray}
Z(\lambda, N) &:=& \int\, d \mu (H )  \, e^{- N  S_0(\lambda, H)} , \label{ZH}
\\
F(\lambda, N) &:=& \frac{1}{N^2}\log Z(\lambda, N) . \label{FH} 
\end{eqnarray}
We perform the one-to-one  change of variables 
(not singular for $\lambda $ real positive)
\bee
K:= H \sqrt{1 +\lambda H^{2p-2}}, \quad K^2= H^2 +\lambda H^{2p} ,  \label{change}
\ee
and put $T_p := \frac{H^2}{K^2}$. The corresponding Fuss-Catalan equation is \cite{Krajewski:2017thd}:
\begin{equation}
z T_p^p(z) - T_p(z) + 1 = 0\, ,
\label{FCEq}
\end{equation}
with $z := -\lambda K^{2p-2}$. The change of variables inverts to $H(K):= K \sqrt{T_p (z)}$. We keep implicit that $H(K)$ also depends on $\lambda$ and also write simply $H$ for $H(K)$ and so on
when no confusion is expected. We also define the corresponding scalar functions
\bea f_\lambda (u) := \sqrt{T_p (-\lambda u^{2p-2})}, \;\;
h_\lambda (u) := uf_\lambda (u) , \;\;
k_\lambda (v) := v \sqrt{ 1 + \lambda v^{2p-2}} .\label{definitionhk:eq}
\eea
They will be used below to express $H$ in terms of $K$ and $K$ in terms of $H$, as $h_\lambda$
and $k_\lambda$ are inverse of each other \bee 
h_\lambda \circ k_\lambda (z) = k_\lambda \circ h_\lambda (z)=z
\ee
in the cut complex plane which is the natural domain of the square root and Fuss-Catalan functions \cite{Rivasseau:2007fr}.
The \emph{Jacobian} of the change of variables \eqref{change} produces a new non-polynomial interaction.
According to \cite{Krajewski:2017thd}
it writes
\begin{eqnarray}
\Big|\frac{\delta  H}{\delta K}\Big| = \Big|\det \frac{ H\otimes \mathbf{1} - \mathbf{1} \otimes H}
{K \otimes \mathbf{1} - \mathbf{1} \otimes K}\Big|\,.
\label{JH}
\end{eqnarray}
In the following we do not take the absolute of $\frac{\delta  H}{\delta K}$, since it is positive for $\lambda>0$
and can be extended to all other $\lambda$'s from the pacman domain by means of the analytical continuation.
We prove the positivity of $\frac{\delta  H}{\delta K}$ in Appendix \ref{PosJ}.
Applying to \eqref{JH} the trace-log formula of \cite{Krajewski:2017thd}
we obtain the following
expression for the partition function
\begin{eqnarray}
Z(\lambda, N) &=& \int\, d\mu( K)\, \exp\{\LV(\lambda, K)\}
\,, \\
\LV(\lambda, K)&=& 
\Tr_\otimes \log
\frac{\partial  H}{\partial K} =
\Tr_\otimes \log\frac{H \otimes \mathbf{1} - \mathbf{1} \otimes {H}}{K \otimes \mathbf{1} - \mathbf{1} \otimes K} .
\label{SpH1}
\end{eqnarray}


The application of the LVE machinery goes along the same line as for complex matrices and 
allows one to express the free energy of the Hermitian 
matrix model as a sum over trees.

We first expand the partition function  as
\bea 
Z(\lambda, N) &=& \sum_{n = 0}^\infty \frac{1}{n!} 
  \int\, d\mu (K)\,\prod_{i = 1}^n\, \LV(\lambda, K_i)\,.
\label{LVE2}
\eea
Then we apply the BKAR formula \cite{BK}-\cite{AR1} as in \cite{Krajewski:2017thd}.
It replaces the covariance $C_{ij} = N^{-1}$ by $C_{ij}(x) = N^{-1}x_{ij}$ ($x_{ij} = x_{ji}$) evaluated at $x_{ij} = 1$
for $i \neq j$ and $C_{ii}(x) = N^{-1} \; \forall i$ and expands according to the BKAR forest Taylor formula. The result is 
a sum over the set $\gF_n$
of forests $\cF$ on $n$ labeled vertices
\bea 
Z(\lambda, N) &=& \sum_{n = 0}^\infty \frac{1}{n!} \;\; \sum_{\cF\in \gF_n} \;
\int dw_\cF\  \partial_\cF  {\cal Z}_n\;  \Big|_{x_{ij} = x_{ij}^\cF (w)}\\
\label{LVE3}
\int dw_\cF &:= & \prod_{(i, j) \in \cF} \int_0^1dw_{ij} \; ,  \quad
\partial_\cF  := \prod_{(i, j) \in \cF} \frac{\partial}{\partial x_{ij}} \; ,\label{LVEH5}\\
{\cal Z}_n &:=& \int\, d\mu_{C(x)}(\{K\}) \,\prod_{i = 1}^n\,  \LV(\lambda, K_i)
\label{LVEzn}
\\
x_{ij}^\cF &:=& \left\{ 
\begin{array}{c}
\hspace{-1.4cm}\text{inf}_{(k,l)\in P^\cF_{i\leftrightarrow j}} w_{kl}~~~~~~~~~~~ \text{if}~ P^\cF_{i\leftrightarrow j}~ \text{exists}\,, \\ 
0~~~~~~~~~~~~~~~~~~~~~~~~~~~~\text{if}~ P^\cF_{i\leftrightarrow j}~ 
\text{does not exist}\, .
\end{array}
\right.
\label{LVE4a}
\eea
In this formula $w_{ij}$ is the weakening parameter of the edge $(i,j)$ of the forest, 
and $P^\cF_{i\leftrightarrow j}$ is the unique path in $\cF$ joining $i$ and $j$ when it exists  \cite{BK}-\cite{AR1}.

The differentiation with respect to $x_{ij}$ in (\ref{LVEH5}) results in
\begin{equation}
\frac{\partial}{\partial x_{ij}}\Big(\int\, d\mu_{C(x)}(\{K\})  \, f(K)\Big) 
= \frac{1}{N} \int\,  d\mu_{C(x)}(\{K\})  \Tr\big[\frac{\partial}{\partial K_i}\frac{\partial}{\partial K_j}\big] f(K)\,.
\end{equation}

The operator $\Tr\big[\frac{\partial}{\partial K_i}\frac{\partial}{\partial K_j}\big]$ acts on two distinct loop vertices ($i$ and $j$) and connects them by an  edge.
Introducing the condensed notations 
\bee
\partial_\cF^K = 
 \prod_{(i, j) \in \cF} \Tr\big[\frac{\partial}{\partial K^\dagger_i}\frac{\partial}{\partial K_j}\big],\quad
 \LV_n =\prod_{i = 1}^n\,  \LV(\lambda, K_i) ,
\ee
we obtain
\bee 
Z(\lambda, N) = \sum_{n = 0}^\infty \frac{1}{n!} \, \sum_{\cF\in \gF_n} N^{-\vert \cF \vert}  \int dw_\cF 
 \int  d\mu_{C(x)}(\{K\})  \partial_\cF^K   \LV_n
 \Big|_{x_{ij} = x_{ij}^\cF (w)} .
  \label{LVE4}
\ee

As usual, since the right hand side of \eqref{LVE4}
is now factorized over the connected components of the forest $\cF$, which are spanning trees, 
its logarithm, which selects only the connected parts, is expressed by \emph{exactly the same formula} but summed
over \emph{trees}. For a tree on $n$ vertices we have $\vert \cT \vert  = n-1$ and taking into account
the $N^{-2}$ factor in the normalization of $F$ in \eqref{FH} we obtain the 
expansion of the free energy as (remark the sum which starts now at $n=1$ instead of $n=0$)
\bea 
F(\lambda , N) &=& \sum_{n = 1}^\infty \frac{1}{n!} \, \sum_{\cT\in \gT_n} A_\cT , \label{LVE6}
\\
A_\cT &:=&  N^{-n-1}   \int dw_\cT 
 \int  d\mu_{C(x)}(\{K\})  \partial_\cT^K   \LV_n
\Big|_{x_{ij} = x_{ij}^\cT (w)},
\label{LVEtree}
\eea
where $\gT_n$ is the set of spanning trees over $n\ge 1$ labeled vertices.

\begin{figure}[!ht]
\begin{center}
{\includegraphics[width=5cm]{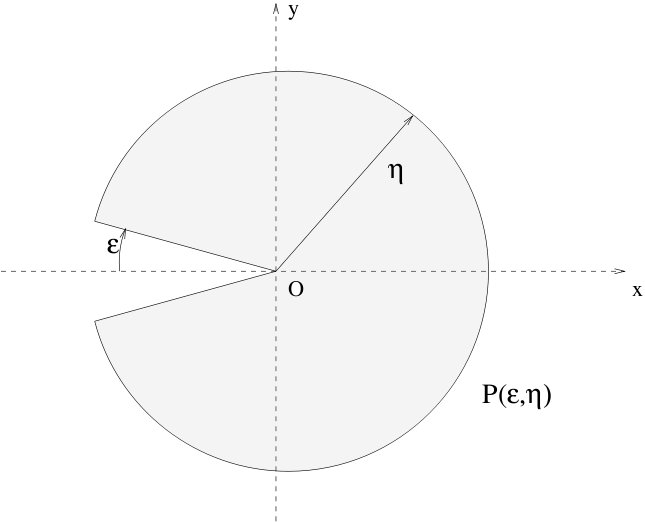}}
\end{center}
\caption{A pacman domain.}
\label{pacman}
\end{figure}

From now on let us write $O(1)$
for a generic constant (independent of $N$) which however may depend on $\epsilon$.
Our main result for the Hermitian matrix model is given by the following theorem, similar
to the complex case of \cite{Krajewski:2017thd}
\begin{theorem} \label{mainH} For any $\epsilon>0$ there exists $\eta$ small enough such that
the expansion \eqref{LVE6} is absolutely convergent and defines an analytic function of $\lambda$,
uniformly bounded in $N$, in the \emph{uniform in $N$} ``pacman domain" 
\bee
P(\epsilon, \eta) := \{0< |\lambda |<\eta, \vert \arg \lambda \vert < \pi - \epsilon\}\,.
\ee 
More precisely, for fixed $\epsilon$ and $\eta$ as above there exists a constant $O(1)$, independent of $N$ such that 
for $\lambda \in P(\epsilon, \eta)$
\bee   \sum_{n = 1}^\infty \frac{1}{n!} \, \sum_{\cT\in \gT_n}  \vert A_\cT \vert \le O(1) < \infty \label{unifobouH}\,.
\ee
\end{theorem}

\section{Proof of Theorem \ref{mainH}}
\label{secder}

We need first to compute $ \partial_{\cT^K}   \LV_n$ assuming $n \ge 2$ (as usual the special case $n=1$
requires an additional integration by parts).
Since trees have arbitrary coordination numbers we need a formula for the action on a vertex factor $\LV (\lambda, K_i) $ of a certain number $r^i \ge 1$ of
derivatives $\frac{\partial}{\partial K_i}$ with $\sum_i r^i = 2n-2$. 

Let us fix a given loop vertex and forget for a moment to write the vertex index $i$. We need to develop a formula for the action of 
a product of $r$ $\frac{\partial}{\partial K}$ 
derivatives on $\LV$. To perform this computation we use  $r$  $\sqcup$ symbols as in \cite{Krajewski:2017thd} to 
indicate the $r$ pairs of external indices of the 
$r $ $\frac{\partial}{\partial K}$ 
derivatives. The final tree amplitude will be obtained later by gluing these $\sqcup$ symbols along the edges of the trees.

The first $\frac{\partial}{\partial K}$ derivative 
is a bit special as it destroys forever the logarithm in $\LV$ and gives
\bee \bigg[\frac{\partial}{\partial K}\bigg]  \Tr_\otimes \log \big[\bbone_\otimes +\Sigma  \big] = 
\big[\bbone_\otimes +\Sigma \big]^{-1} 
\frac{\partial \Sigma }{\partial K} . \label{firstder}
\ee
\begin{figure}[!ht]
\begin{center}
{\includegraphics[width=12cm]{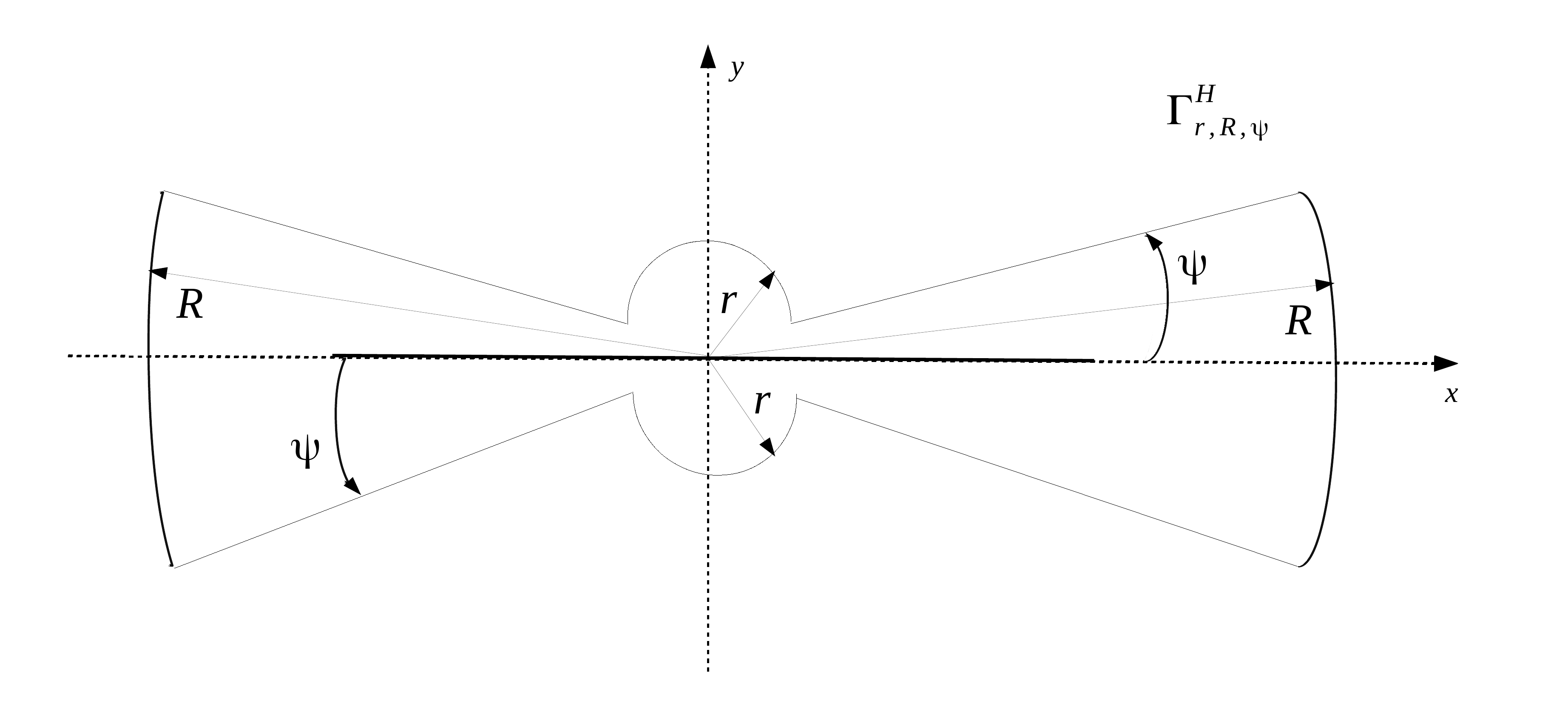}}
\end{center}
\caption{A keyhole contour $\Gamma$ encircling the spectrum of $H$
which for $H$ Hermitian 
lies on a real axis positive segment like the one shown in boldface.}
\label{keyholeencirc1}
\end{figure}
We can use holomorphic functional matrix calculus as in \cite{Krajewski:2017thd}
to  write 
\bea \big[\bbone_\otimes 
+\Sigma_\lambda(K) \big]^{-1} &=& 
\frac{\delta  K}{\delta H}
= \frac{ K\otimes \mathbf{1} - \mathbf{1} \otimes K}
{H \otimes \mathbf{1} - \mathbf{1} \otimes H},\label{sigmadef}\\
\Sigma_\lambda (K)  &:=&  \oint_{\Gamma} du\; [h_\lambda(u)-u]  \frac{1}{u-K}  \otimes  \frac{1}{u-K},
\label{sigmadef1}
\eea
where the contour $\Gamma$
is any  contour enclosing 
the spectrum of $K$.
This spectrum lies on the real axis so the contour has to enclose this real axis, avoiding
any singularity of the function $h_\lambda$. 
This function $h_{\lambda}(u)$ is analytic on the complex plane with $2p-2$ cuts at $-\lambda u^{2p-2}\in\big[\frac{(p-1)^{p-1}}{p^{p}},+\infty\big]$,
hence such that
\bee
|u|\geq |\lambda|^{-\frac{1}{2p-2}}
\frac{(p-1)^{1/2}}{p^{p/(2p-2)}}\; , \quad
\text{arg}(u)=\frac{\pi-\text{arg}(\lambda)}{2p-2}+\frac{k\pi}{p-1}\; ,
\ee
for  $k=-p+2,\dots,p-1$. There are plenty of possible choices for
$\Gamma$ to avoid the cut, but one of the simplest, inspired by \cite{Krajewski:2017thd} is to choose for $\Gamma$
the \emph{finite} symmetric ``keyhole" parametrized by $R,r$ and $\psi$
as in Figure \ref{keyholeencirc1}, with $R$ large (e g larger than $2\Vert K \Vert$), $r =1$ and $\psi$
small  as $\epsilon \to 0$ (eg $\psi = \epsilon/2$).
Noticing that $g_\lambda (u)= [h_\lambda(u)-u]$ vanishes at $\lambda =0$, it can be written as
\bee g_\lambda (u) = \int_0^\lambda dt \partial_tg_t (u) = - \frac{1}{2}\int_0^\lambda dt u^{2p-1} e_t (u) f_t (u), \label{deffuncg}
\ee
where we define as in \cite{Rivasseau:2017hpg} 
$e_t (u):=\frac{T'_p}{T_p} (-t u^{2p-2})$.

From now on we use $O(1)$ as a generic name for any inessential numerical constant (which may depend on the parameters 
$\epsilon$ and $\eta$
of our fixed pacman domain).
\begin{lemma}
On the contour $\Gamma$ we have the bound
\bee \vert g_\lambda (u)\vert \le
O(1) \vert \lambda\vert^{\frac{1}{4p^2}}
\vert u\vert ^{1+ \frac{1}{2p}-\frac{1}{2p^2}}.
\label{gbound}
\ee
\end{lemma}
\proof We can use the rather standard estimates on $T_p$ and $E_p(z)=\frac{T'_p}{T_p}(z)$ proven in section III of \cite{Rivasseau:2017hpg} (see Lemma III.1). In particular it is proven there that in a domain avoiding a small angular opening $\epsilon$ around the cut of $T_p$ we have
\bee
T_p (z) \le (1 + \vert z \vert )^{-\frac{1}{p}} , \quad E_p(z)\le \frac{O(1)}{(1 + \vert z \vert )}. \label{recallE}
\ee
In our case this means that on our contour $\Gamma$, 
for any $0 <\delta <1$ there is a 
constant $C_\delta$ such that 
\bee  \vert e_t f_t (u) \vert 
\le  
\frac{C_\delta}{[\vert t\vert  \vert u\vert^{2p-2}]^{(1+\frac{1}{2p})(1-\delta)}},
\ee
Choosing $\delta= \frac{1}{2p}$
gives \eqref{gbound}.
\qed

From now on and when there is no risk of ambiguity we write simply $\Sigma$ for $\Sigma_\lambda (K)$.
It remains to compute $\frac{\partial \Sigma}{\partial K}$ in \eqref{firstder}.
The $\partial $ derivative can act on the left or right side of the tensor product so that
\bea
\frac{\partial \Sigma}{\partial K}&=&  
\oint_{\Gamma} du\; g_\lambda(u)\biggl[\frac{1}{u-K} \sqcup\frac{1}{u-K}\otimes \frac{1}{u-K}
\nonumber \\&+&  
\frac{1}{u-K} \otimes  \frac{1}{u-K}\sqcup\frac{1}{u-K} \biggr]. \label{derivsigma}
\eea

Then the next derivatives iterate in a similar pattern.
Each $\frac{\partial}{\partial K}$ derivative
\begin{itemize}
\item
either derives a
$[1 + \Sigma]^{-1}$ factor and creates a new 
$\frac{\partial \Sigma}{\partial K}$ through the resolvent formula (easily checked algebraically)
\bee \frac{\partial  }{\partial K}[1 + \Sigma]^{-1}= -
[1 + \Sigma]^{-1}\frac{\partial \Sigma  }{\partial K}[1 + \Sigma]^{-1} .
\ee
To this $\frac{\partial \Sigma  }{\partial K}$ is associated a new integration contour through \eqref{derivsigma}.
\item or derives again an existing $\Sigma$. In this case it results in no new contour but in a new $\sqcup$ and a multiplication by a new factor $\frac{1}{u-K} \otimes \bbone + \bbone \otimes \frac{1}{u-K}$.
\end{itemize}
The combinatorics to sum over the choices is the usual one
relying on the Fa\`a di Bruno formula. Since it
is similar to the one explained in detail of the complex case \cite{Krajewski:2017thd}  we wont discuss it further here.
The result for a loop vertex of degree $r$, hence with $r$
``corners" $c$ between half edges,
is a sum over sequences of $1 \le m \le r$
contour-corner operators $O^c$
and $r-m$ derivative-corner operators $\tilde O^c$. Each such operator is sandwiched between two $\sqcup$ insertions. 
The tree amplitude $A_\cT$, as also detailed in \cite{Krajewski:2017thd}, is obtained by identifying the two ends of each pair of $\sqcup$ symbols along each edge of $\cT$. This pairing of the $2n-2$ $\sqcup$ symbols then exactly glue the $2n$ traces of the tensor products present in the $n$ vertices into $n+1$ traces.

\begin{figure}[!ht]
\begin{center}
{\includegraphics[width=9cm]{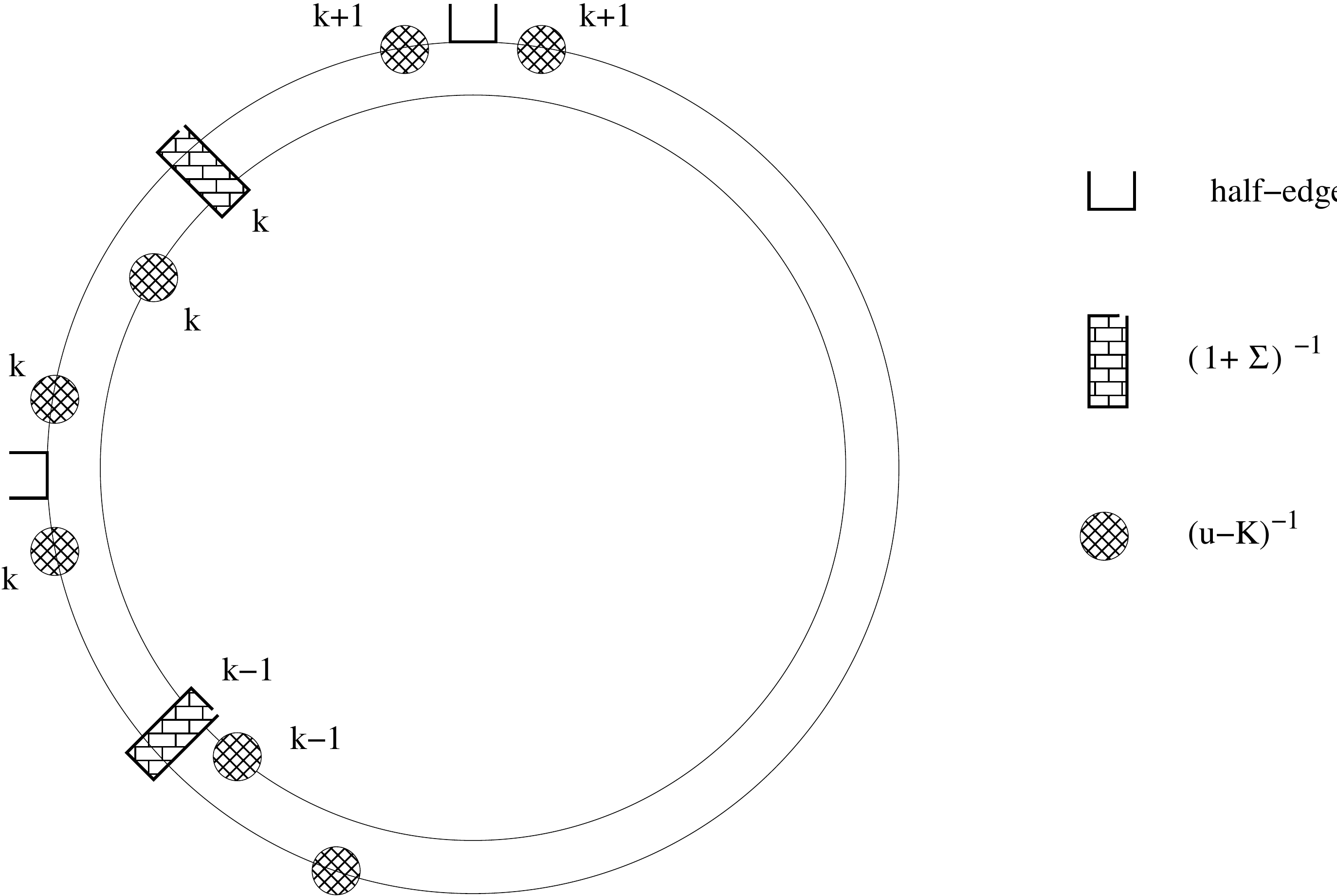}}
\end{center}
\caption{A vertex with some of its corner operators. The label $k$ indicates the corresponding contour variable. The upper left corner between the two half-edges $\sqcup$ symbols contains three $(u-K)^{-1}$ operators with indices $k$, $k$ and $k+1$.}
\label{corneropfig}
\end{figure}

However we have not yet given the exact formula for the
contour-corner operators $O^c$ and the derivative-corner operators $\tilde O^c$. The derivative-corner operators are simply defined as
\bee \tilde O^c (u_{k_c}) := \frac{1}{u_{k_c}-K} \otimes \bbone + \bbone \otimes \frac{1}{u_{k_c}-K}
\ee
where $k_c \in [1, m]$ is the index of its ``parent" contour-corner. For contour-corner operators the formula is more interesting and contains a subtlety, pictured in Figure \ref{corneropfig}. To each contour corner $c_k$, $k \in [1, m]$ is associated a contour integral $\int d u_k g_\lambda(u_k)[1 + \Sigma]^{-1}$ factor. But since the two 
$\frac{1}{u_k-K}$ operators of the same side of the $\otimes$ tensor product in \eqref{derivsigma} are separated by a $\sqcup$ they do not belong to the same \emph{corner}. Therefore we must attribute one of them to the next corner, in a cyclic way around the vertex.
As a consequence the formula for a contour-corner operator $O^{c_k}$, $k \in [1, m]$ contains both $u_k$ and $u_{k+1}$ (with the cyclic convention $u_{m+1} :=u_1$). 
Taking out in front of the loop vertex the global  contour integral  $\prod_{k=1}^m \oint_{\Gamma} du_k g_\lambda(u_k)$ we define the $k$-th contour-corner operator therefore as
\bea O^{c_k}(u_k, u_{k+1})&:=& \big[\bbone_\otimes +\Sigma \big]^{-1}  \bigg[  \frac{1}{u_k-K}\frac{1}{u_{k+1}-K}\otimes  \frac{1}{u_k-K} \nonumber \\
&+&\frac{1}{u_k-K}\otimes  \frac{1}{u_k-K}\frac{1}{u_{k+1}-K}\bigg] .
\eea
Remark that this operator contains therefore 
both $u_k$ and $u_{k+1}$.
 
To bound the amplitude $A_\cT$ we first have to bound the corner operators. The bound on a derivative-corner operator is rather trivial. Since the contour $\Gamma$ is never closer than $r \sin \psi$ to 
the spectrum of $K$ (see Figure \ref{keyholeencirc1}) we have
\bee \Vert  \tilde O^c (u_{k_c}) \Vert \le 2(r\sin \psi )^{-1}. 
\ee
For the contour-corner operators the bound is more delicate.
We remark that all corner operators of a given loop vertex  commute as they involve to the same replica field $K$. In fact they are diagonalized by the tensor basis $e_i \otimes e_j$ where $e_i$ is the basis diagonalizing $K$. Let us call $\mu_i$ the eigenvalue of $K$ on $e_i$.
 The operator $O^{c_k} (u_k, u_{k+1})$ is diagonal on the basis $e_i \otimes e_j$, with eigenvalues
\bea O_{ij}^{c_k}(u_k, u_{k+1}) &=& \big[\bbone_\otimes +\Sigma \big]_{ij}^{-1} 
 \big[  \frac{1}{u_k-\mu_i}\frac{1}{u_{k+1}-\mu_i}\otimes  \frac{1}{u_k-\mu_j}\nonumber \\
&+&\frac{1}{u_k-\mu_i}\otimes  \frac{1}{u_k-\mu_j}\frac{1}{u_{k+1}-\mu_j}\bigg]
 \label{keycont2} .
\eea

\begin{lemma}\label{lemma3.1}
For complex $\lambda$ such that $|\text{arg}(\lambda)|\leq\pi-\epsilon$
there exists some constant $O(1)$ such that
\bea
\vert(\bbone_\otimes +\Sigma )_{ij}^{-1}\vert&\leq& O(1) \Lambda_{ij}\label{Rbound}
\\
\Lambda_{ij}&:=& \sup \{1, \vert \lambda \vert^{\frac{1}{2p}} |\mu_{i}|^{1-\frac{1}{p}},
 \vert \lambda \vert^{\frac{1}{2p}} |\mu_{j}|^{1-\frac{1}{p}} \}.\label{Lambdabound}
 \eea
\end{lemma}

\prf
Calling $\nu_i = h_\lambda (\mu_i)$, \eqref{sigmadef} means that 
\bee (\bbone_\otimes +\Sigma )_{ij}^{-1}= \frac{k_\lambda (\nu_i) -k_\lambda (\nu_j )}{\nu_i - \nu_j}.
\ee
hence it is bounded by $\sup_{\nu \in [\nu_i, \nu_j]} \vert k'_\lambda (\nu) \vert $ where the sup is taken along the $[\nu_i, \nu_j]$ segment. $k'_\lambda$ can be explicitly computed (see \eqref{definitionhk:eq}) and from the large $z$ behaviour of the function $T (z)\sim z^{-1/p}$  derived from its functional equation \eqref{FCEq} the bound follows easily on the pacman domain.
\qed
\medskip

The next step is to compensate the growth of this bound as $\mu_i$ or $\mu_j$ becomes large with the decay hidden in the $\frac{1}{u - \mu}$ factors. 
The contour $\Gamma$ in Figure \ref{keyholeencirc1}  has been chosen so that
everywhere along the contour
\bee
\bigg|\frac{1}{u-\mu}\bigg|\leq O(1) \inf 
(\frac{1}{1+ \vert u  \vert }, \frac{1}{1+ \vert\mu\vert}).
\label{contbou}
\ee
for some other constant $O(1)$. 
\begin{lemma}\label{lemma3.2}
For complex $\lambda$ such that $|\text{arg}(\lambda)|\leq\pi-\epsilon$
\bee \Vert O^{c_k}(u_k, u_{k+1})\Vert \le O(1) \frac{1}{(1 + \vert u_k\vert )^{1+ \frac{1}{p}}} \frac{1}{1+ \vert u_{k+1}  \vert } .
 \label{keycont4}
\ee
\end{lemma}
\proof
Suppose eg $\Lambda_{ij} =\vert \lambda^{\frac{1}{2p}} \vert \vert \mu_i \vert^{1-\frac{1}{p}}$. Using \eqref{contbou} we bound the $\frac{1}{u_k-\mu_i}$
factor in \eqref{keycont2} as
\bee \bigg\vert \frac{1}{u_k-\mu_i}\bigg\vert \le 
\biggl[\frac{1}{1+ \vert u_k  \vert }\biggr]^{1/p} \biggl[\frac{1}{1+ \vert\mu_i\vert}\biggr]^{1 - 1/p} .
\ee
Combining with \eqref{Rbound} leads to
\bee \vert O_{ij}^{c_k}(u_k, u_{k+1})\vert \le O(1) \biggl[\frac{1}{1 + \vert u_k\vert }\biggr]^{1 + 1/p} \frac{1}{1+ \vert u_{k+1}  \vert } .
\label{keycont3}
\ee
The other cases $\Lambda_{ij} =\vert \lambda^{\frac{1}{2p}} \vert \vert \mu_j \vert^{1-\frac{1}{p}}$
or $\Lambda_{ij} = 1$ are obviously similar. Since the bound 
\eqref{keycont3} is independent of $i$ and $j$, it implies
\eqref{keycont4}, with $C_{\epsilon,2}=2C_\epsilon [C_{\epsilon,1}]^3$.
\qed

\medskip
Still keeping the integral over the contour parameters
for later we now glue the $\sqcup$ operators and perform all traces. We obtain
\begin{lemma}\label{lemma3.4}
There exists some constant $O(1)$ such that
\bee \vert A_\cT \vert \le [O(1)]^n \prod_v \prod_{k=1}^{m(v)}\oint_\Gamma \vert g_\lambda(u_k)\vert  \biggl[\frac{1 }{1+ \vert u_k \vert}\biggr]^{2+ \frac{1}{p}}du_k .\label{treebou1}
\ee
\end{lemma}
 \proof
We bound recursively all tree traces.  The simplest way to understand how it works is to start from a leaf $f$, which has $r=m=1$. The associated operator is therefore a single contour-corner operator $O^c$ whose norm, by \eqref{keycont4}, is bounded by $O(1) [\frac{1 }{1+ \vert u_k \vert}]^{2+ \frac{1}{p}} $. The amplitude for $A_\cT$ contains a partial trace on one $\cH$ factor of the tensor product $\cH \otimes \cH$ of the leaf vertex, leading to a simpler operator on $\cH$ only, with norm bounded by $N O(1) [\frac{1 }{1+ \vert u_k \vert}]^{2+ \frac{1}{p}}$. After gluing this factor between the two appropriate corners in the parent vertex $v(f)$ we can find a new leaf and iterate. This leads to in leads to the bound. Indeed this induction collects 
exactly $n+1$ factors $N$ (since the last vertex of the tree brings two such factors). This exactly compensates with the $N^{-n-1}$ factor in \eqref{LVEtree}. Finally the 
$ \int dw_\cT  \int d\mu_{C(x)}(\{K\})$ integrals are normalized so do not add anything to the bounds.
\qed
 
To complete the bound on 
$A_\cT $ it remains only to perform all contour integrals. From the choice of our contour
\begin{lemma}
There exists some constant $O(1)$ such that
\bee \oint_\Gamma \vert g_\lambda(u)\vert  \biggl[\frac{1 }{1+ \vert u \vert}\biggr]^{2+ \frac{1}{p}}du\le O(1) \vert \lambda \vert^{\frac{1}{4p^2}} . \label{lasbou}
\ee
\end{lemma}
\proof
Inserting \eqref{gbound} proves 
\eqref{lasbou}
 since the integral $\oint_\Gamma \frac{\vert u \vert^{1+\frac{1}{2p}- \frac{1}{2p^2}} }{(1+ \vert u \vert)^{2+ \frac{1}{p}}} du$ is absolutely convergent and bounded by a constant at fixed $p$. 
\qed

\medskip
Finally since each vertex has at least one contour operator, the number of $\vert \lambda \vert^{\frac{1}{4p^2}}$ factors in the bound is at least $n$. 
Taking into account that the number of (labeled) trees is bounded by $K^n n!$ for some constant $K$
completes the proof of \eqref{unifobouH}, hence of Theorem \ref{mainH}.

\section{The (not-so-)trivial $n=1$ tree}
\label{trivialtreeH}

This section is devoted to establish a not-so trivial bound
on the trivial tree amplitude with a single vertex, namely
\bee
A_{\cT_\emptyset}= N^{-2} \int\, d\mu \   \LV (\lambda, K), \quad  \LV (\lambda, K)=\Tr_\otimes \log  \frac{\partial H}{\partial K}.
\ee
More precisely it is devoted to prove
\begin{lemma}
\label{Alemma}
We have 
\bee \vert  A_{\cT_\emptyset}\vert  \le O(1) \vert \lambda\vert^{\frac{1}{2p(2p-2)}}\label{Alemmabou}
\ee
\end{lemma}
\proof
Let us rewrite the Jacobian matrix as
\begin{eqnarray}
  \frac{\partial H}{\partial K} &=& 
 {\bf 1}_\otimes +  \Sigma_\lambda (K) = {\bf 1} \otimes f_\lambda (K) + K  \frac{\partial f_\lambda (K)}{\partial K} \\
  &=& 1\otimes f_\lambda(K) + \oint_{\Gamma} du
  f_\lambda(u)\frac{K}{u - K}\otimes\frac{1}{u - K}\end{eqnarray}
so that $A_{\cT_\emptyset} =  A_1
+ A_2$ with
\begin{eqnarray}
A_1 &:=&  \frac{1}{2N}\int d \mu \;  \Tr \log T_p (- \lambda K^{2p-2} ) \label{bada1}\\
&=&-\frac{1}{2N}\int d \mu \;  \Tr
\int_0^\lambda dt  \oint_{\Gamma} 
u^{2p-3} e_t (u)
\frac{K}{u - K} du,\label{bettera1}
\eea
and
\bea
A_2 &:=& N^{-2}\int d \mu \; \Tr_\otimes \log \big[ {\bf 1}_\otimes + V_\lambda (K) \big],\\
{\bf 1}_\otimes + V_\lambda (K) 
&=&\big[ 1 \otimes f_\lambda (K)\big]^{-1} [{\bf 1}_\otimes + \Sigma_\lambda (K) \big].\label{def1+V}
\eea
We can write $V_\lambda(K)$ 
as a double contour integral
\bea
V_\lambda(K) &:=& \oint_{\Gamma} du \oint_{\Gamma'} dv  \, \phi (\lambda, u,v)
\frac{K}{u - K}\otimes\frac{1}{v - K},
\\
\phi (\lambda, u,v) &:=&\frac{1}{u-v}  \frac{f_\lambda(u)}{f_\lambda(v)},
\eea
where $\Gamma'$ is another keyhole contour similar to $\Gamma$ surrounding the spectrum of $K$
but inside $\Gamma$ and with half its opening angle.

The $A_1$ part is easy to bound and to prove in addition that it tends to zero as $\lambda \to 0$. We simply integrate by parts the $K$ numerator in \eqref{bettera1} to get
\bee
A_1=-\frac{1}{2N^2}\int d \mu \;  \Tr_\otimes
\int_0^\lambda dt  \oint_{\Gamma} du
u^{2p-3} e_t (u)
\frac{1}{u - K}\otimes \frac{1}{u - K}.\label{betterra1}
\ee
and recalling \eqref{recallE} one
can use $\vert e_t (u)\vert \le [\vert t\vert \vert u\vert ^{2p-2}]^{-1 + \frac{1}{2p-2}}$
to conclude easily that $A_1$
is $O(1) \vert \lambda\vert^{\frac{1}{2p-2}}$.

Turning to $A_2$ we remark that
$V$ vanishes at $\lambda = 0$, hence we can rewrite it as $\int_0^\lambda dt\; \partial_t V_t$ with
\bea
 \partial_t V_t (K)&=&\oint_{\Gamma} du \oint_{\Gamma'} dv  \, \partial_t\phi (t, u,v)
\frac{K}{u - K}\otimes\frac{1}{v - K} , \label{numeratnice}
\eea
with $\partial_t\phi$ easily computed as
\bee \partial_t\phi (t, u,v)=
\frac{1}{2(u-v)}\frac{f_t (u)}{f_t (v)}\big[v^{2p-2} e_t (v)-
u^{2p-2}e_t (u)\big]
  ,
\ee
so that
\bea
A_2 &:=& N^{-2}\int d \mu \int_0^\lambda dt\;\oint_{\Gamma} du \oint_{\Gamma'} dv  \, \partial_t\phi (t, u,v) \\
&&
\Tr_\otimes  \big[ {\bf 1}_\otimes + V_\lambda (K) \big]^{-1}
\frac{K}{u - K}\otimes\frac{1}{v - K}
\eea

On the contours we can easily bound
$\frac{f_t(u)}{f_t (v)}$
by $1 + \vert v \vert^{1-\frac{1}{p}} $,
hence $\frac{1}{u-v}\frac{f_t(u)}{f_t (v)}$
by $O(1) [1 + \vert u \vert + \vert v \vert]^{-\frac{1}{p}} $,
$\vert u^{2p-2}e_t(u) \vert $ 
by $\vert t\vert ^{-1 + \frac{1}{2p(2p-2)}}
\vert u \vert^{\frac{1}{2p}} $ and similarly
$\vert v^{2p-2}e_t(v) \vert $ 
by $\vert t\vert ^{-1 + \frac{1}{2p(2p-2)}}\vert v \vert^{\frac{1}{2p}}$, so that
finally
\bee
\vert \partial_t\phi(t,u,v) \vert \le O(1)\frac{ \vert t\vert^{-1 + \frac{1}{2p(2p-2)}}}{[1 +\vert u\vert+ \vert v\vert]^{\frac{1}{2p}}} .
\label{partialphibou}
\ee

Then we integrate by part the $K$ numerator in \eqref{numeratnice}. We get five terms, two of which are ``triple trace" and three of which ``single trace". Since we remain in the commutative algebra generated by $K$ we can diagonalize all tensor products
and compute all traces. We define 
the resolvent ${\bf R}_t (K) := [{\bf 1}_\otimes + V_t(K) \big]^{-1}$ and
write simply $V$ for $V_t(K)$, ${\bf R}$ for ${\bf R}_t(K)$ and so on.
Remember that from \eqref{def1+V}
we have
\bee [{\bf 1}_\otimes + V \big]^{-1}
=\big[ 1 \otimes f\big] \big[{\bf 1}_\otimes + \Sigma \big]^{-1}.
\ee
${\bf R}$ is diagonal on the basis $e_i \otimes e_j$ with eigenvalue ${\bf R}_{ij}$. Since 
$1 \otimes f$ is also diagonal on that basis $e_i \otimes e_j$, with eigenvalue
decaying as $\vert t \mu_j^{2p-2}\vert^{-1/2p}$, from \eqref{Rbound}-\eqref{Lambdabound} we get easily
\bee \vert {\bf R}_{ij}\vert \le O(1)
\sup \{1, \vert t \vert^{\frac{1}{2p}}\vert  \mu_i\vert^{1-\frac{1}{p}} \} .
\label{calRbound}
\ee
\begin{figure}[!ht]
\begin{center}
{\includegraphics[width=12cm]{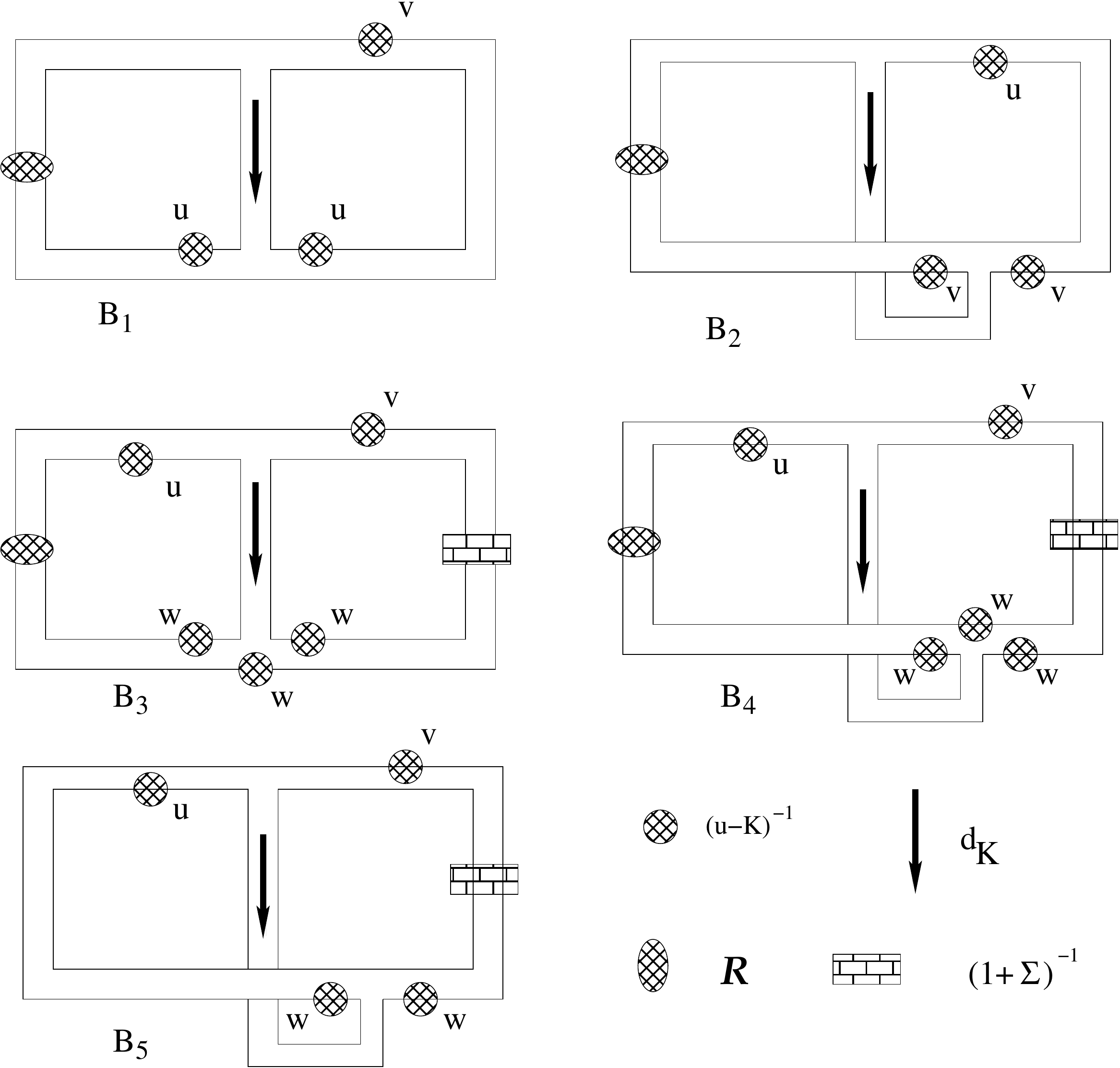}}
\end{center}
\caption{The five terms $B_1$, $B_2$, $B_3$, $B_4$ and $B_5$. The arrow indicates the action of the $\partial_K$ matrix derivative.}
\label{bterms}
\end{figure}
We define ${\bf R}_{diag}$ and $({\bf 1}_\otimes + \Sigma )^{-1}_{diag}$
as the diagonal ``single thread"  $N$ by $N$
matrix with eigenvalue
$ {\bf R}_{diag}^i:= {\bf R}_{ii}$ or  $({\bf 1}_\otimes + \Sigma )^{-1}_{ii}$
on $e_i$, and perform a careful analysis of the tensor threads involved,
hopefully helped by Figure \ref{bterms}. It gives 
\bee
A_2 = N^{-3}\int d \mu \; \int_0^\lambda dt  \oint_{\Gamma} du \oint_{\Gamma'} dv  \, \partial_t\phi (t, u,v) \big[ B_1 + B_2 +B_3 + B_4 +B_5 \big],
\ee
where the first two terms are obtained when $\partial_K$ hits $\big[\frac{1}{u - K}\otimes\frac{1}{v - K} \big]$, giving
\bea
B_1&=& 
\Tr_{\otimes^3} \bigg[ \big[ {\bf R} \otimes 1\big] \big[\frac{1}{u - K}\otimes\frac{1}{v - K} 
\otimes \frac{1}{u - K}\big] \bigg],\label{b1}\\
B_2&=& 
\Tr\; {\bf R}_{diag}  \frac{1}{(u - K)(v - K)^2}, \label{b2}
\eea
The last three terms $B_3$, $B_4$ and $B_5$ come from $\partial_K$ hitting ${\bf R}$
\bea\partial_K {\bf R} &=& \partial_K (1 \otimes f)(1+ \Sigma)^{-1} 
\\
&=&  
- {\bf R} \;
[\partial_K \Sigma ] \; (1+ \Sigma)^{-1} + [1 \otimes \partial_K f ] (1+ \Sigma)^{-1}.
\eea
We can recompute  $\partial_K \Sigma$,
as
\bea
\frac{\partial \Sigma}{\partial K}&=&  
\oint_{\Gamma} dw\; w f_t(w)\biggl[\frac{1}{w-K} \sqcup\frac{1}{w-K}\otimes \frac{1}{w-K}
\nonumber \\&+&  
\frac{1}{w-K} \otimes  \frac{1}{w-K}\sqcup\frac{1}{w-K} \biggr], \label{derivsigma1}
\eea
recalling \eqref{derivsigma} but modifying it slightly. 
We also compute easily
\bee \partial_K f = \oint_{\Gamma} dw
f_\lambda (w)\frac{1}{w-K} \otimes \frac{1}{w-K}.
\ee
We obtain
\bea
B_3&=& 
\oint_{\Gamma} dw 
\Tr_{\otimes^3} \bigg[   [ {\bf R}\otimes 1 ][ 1 \otimes  (1+ \Sigma)^{-1}]\nonumber\\
&& \frac{wf_t (w)}{(u - K)(w-K)} \otimes \frac{1}{(v - K)(w-K)}\otimes \frac{1}{w - K} \bigg],\label{b3}\\
B_4&=& \oint_{\Gamma} dw  \;\Tr \;(1+ \Sigma)^{-1}_{diag} {\bf R}_{diag}
\frac{wf_t (w)}{(u - K)(v-K)(w-K)^3},\label{b4}
\\
B_5&=& \oint_{\Gamma} dw \; \Tr \; (1+ \Sigma)^{-1}_{diag} \frac{ f_t (w) }{(u - K)(v-K)(w-K)^2}.
\label{b5}
\eea
\begin{lemma}\label{lastlemma}
We have 
\bee  \sup_{i=1}^5\{\vert B_i \vert\} \le O(1) \frac{1}{1+ \vert u\vert }\frac{1}{1+ \vert v\vert }
\ee
\end{lemma}
\proof

We remark that the decay of $f_t (w)$ at large $w$ means that 
\bea \vert f_t (w)\vert &\le& O(1)\inf \{1  ,
\vert t\vert^{-1/2p} \vert w \vert^{-1+1/p} \} , \label{fdecay} \\
\vert w f_t (w)\vert &\le& O(1)
\inf \{\vert w \vert  ,
\vert t\vert^{-1/2p} \vert w \vert^{1/p} \} 
\label{wfdecay}
\eea

\begin{itemize}
    \item For $B_1$ it is essential to remark that \eqref{calRbound} implies that the growth of ${\bf R}$  can occur only in its \emph{left eigenvalue}. It can therefore be bounded by $O(1)$ using the corresponding left $\frac{1}{u-K}$ factor. There remains 
a $\Vert \frac{1}{u-K}\frac{1}{v-K} \Vert \le  O(1) \frac{1}{1+ \vert u\vert }\frac{1}{1+ \vert v\vert }$ factor.

\item For $B_2$ the growth of ${\bf R}_{diag}$  can be bounded by $O(1)$ using one $\frac{1}{v-K}$ factor. There remains 
a $\Vert \frac{1}{u-K}\frac{1}{v-K} \Vert \le  O(1) \frac{1}{1+ \vert u\vert }\frac{1}{1+ \vert v\vert }$ factor.

\item For $B_3$ the growth of $[{\bf R}\otimes 1 ][1\otimes (1+ \Sigma)^{-1}]$  can be
compensated by the decay of the $\big[\frac{1}{w-K}\big]^{\otimes 3}$ factor. This is again subtle and true only 
because that factor decays separately on 
each of the three
threads in the tensor product, and by \eqref{calRbound} the possible growth of ${\bf R}$ occurs only on the \emph{first thread} and the potential growth of $(1+ \Sigma)^{-1}$ occurs only on the \emph{second or third thread}. Hence they never \emph{conspire on the same thread}.
Using this fact we find the bound
\bea &&\Vert [{\bf R}\otimes 1 ][1\otimes (1+ \Sigma)^{-1}]\frac{1}{w-K}\otimes \frac{1}{w-K}\otimes \frac{1}{w-K}\Vert\nonumber\\
&&
\le O(1)
\sup\{\frac{1}{[1+ \vert w\vert ]^3},
\frac{\vert t\vert^{\frac{1}{2p}} }{[1+ \vert w\vert]^{2 + \frac{1}{p}}}, 
\frac{\vert t\vert^{\frac{1}{p}} }
{[1+ \vert w\vert]^{1 + \frac{2}{p}}}\}.
\eea 
Adding the $wf$ factor combined with bound \eqref{wfdecay}
we get an overall bound on the $B_3$ $w$-integrand
\bee 
O(1)\inf\{ \frac{1}{[1+ \vert w\vert ]^2},
\frac{\vert t\vert^{\frac{1}{2p}} }{[1+ \vert w\vert]^{1 + \frac{1}{p}}}, 
\frac{\vert t\vert^{\frac{1}{p}-\frac{1}{2p}} }
{[1+ \vert w\vert]^{1 + \frac{1}{p}}} \} \le 
O(1)\frac{1}{[1+ \vert w\vert]^{1 + \frac{1}{p}}}
\ee
which is integrable on $\Gamma$, and we still have a $\Vert \frac{1}{u-K}\frac{1}{v-K} \Vert \le  O(1) \frac{1}{1+ \vert u\vert }\frac{1}{1+ \vert v\vert }$ factor left. 

\item For $B_4$ the bound is the same as for $B_3$; just easier because there is no subtle discussion of the threads.

\item For $B_5$ we can simply bound the $f_t (w)$ factor by $O(1)$.
The growth of $(1+ \Sigma)^{-1}_{diag}$ is bounded by one $\Vert \frac{1}{w-K}\Vert^{1- \frac{1}{p}}$ factor.
There remains therefore a factor  $[1+ \vert w\vert]^{-1 - \frac{1}{p}}$ to integrate over $w$, and after that is done,
there remains 
a $\Vert \frac{1}{u-K}\frac{1}{v-K} \Vert \le  O(1) \frac{1}{1+ \vert u\vert }\frac{1}{1+ \vert v\vert }$ factor.
\end{itemize} \qed

Combing Lemma \ref{lastlemma} with the bound \eqref{partialphibou} on 
$\partial_t\phi$, the five $B$ terms are all given by absolutely convergent integrals 
on $u$ and $v$ and the result is that $\vert A_2 \vert \le O(1)\vert \lambda\vert^{\frac{1}{2p(2p-2)}}$.
Combining with the better bound $O(1)\vert \lambda\vert^{\frac{1}{2p-2}}$ on $A_1$ completes the proof of \eqref{Alemmabou}. \qed

Remark that $B_2$, $B_4$ and $B_5$ are proportional to $N^{-2}$, hence subleading
at large $N$.

\appendix 

\section{Appendix: the $O(N)$ real symmetric case}

In the core of this article, we focused on Hermitian matrix models for simplicity. However, the same techniques can be applied to real symmetric and quaternionic Hermitian matrices, with only a few minor changes. In this appendix, we outline how our techniques can be extended to these cases.

First, recall that we work with Hermitian matrices $H\in M_{N}(\mathbb{H})$, $H_{ij}=H_{ji}^{*}$, with $(\mathbb{K}=\mathbb{R},\mathbb{C},\mathbb{H})$. In the real case these are just real symmetric matrices while in the quaternionic case, diagonal elements are real numbers and off-diagonals one form pairs of conjugate quaternions. 
These  are respectively invariant under the groups $\text{O}(N)$, $\text{U}(N)$ and $\text{USp}(N)$ (unitary matrices with quaternionic entries). 

Using the symmetries, the covariance in the normalized Gaussian case is shown to be 
\begin{align}
\int dH\,  H_{ij} H_{kl}\, \exp-\textstyle{\frac{\beta N}{4}}\text{Tr}\,H^{2}= \frac{1}{N}\delta_{il}\delta_{jk}+\frac{2-\beta}{N\beta}\delta_{ik}\delta_{jl},\label{twist}
\end{align}
for complex ($\beta=2$),  real ($\beta=1$) or  quaternionic ($\beta=4$). The first term is conveniently represented by a ribbon and the second one by a twisted ribbon.

In order to derive the general formula for the change of variables, it is convenient to first diagonalize the matrices, $H=U {diag}(\mu_{1},\dots,\mu_{N})U^{\dagger}$, with $\mu_{i}$ real  and $U$ an element of the corresponding unitary group. Let us denote by ${\cal V}_{N}(\mathbb{K})$ the volume of this unitary group after division by diagonal matrices and permutations. Then, the partition function can be written as
\begin{align}
Z&=
\int dH\, \exp-\frac{\beta N}{2}\left\{\frac{1}{2}\text{Tr}\,H^{2}+
\frac{\lambda}{p}\text{Tr}\,H^{2p}
\right\}\\
&={\cal V}_{N}
\int \prod_{i}d\mu_{i}\, 
\prod_{i<j}\big|\mu_{i}-\mu_{j}\big|^{\beta}
\exp-\frac{\beta N}{2}
\sum_{i}
\left\{\frac{1}{2}\mu_{i}^{2}+
\frac{\lambda}{p}\mu_{i}^{2p}
\right\}.
\end{align}
Next, we perform the same change of variable $\mu_{i}=h_{\lambda}(\nu_{i})$ and rewrite the result in terms of a matrix integral over $K$, whose eigenvalues are the $\nu_{i}$'s,
\begin{align}
Z&={\cal V}_{N}
\int \prod_{i}d\nu_{i}\, \big|h'_{\lambda}(\nu_{i})\big|\times\prod_{i<j}\bigg|\frac{h_{\lambda}(\nu_{i})-h_{\lambda}(\nu_{j})}
{\nu_{i}-\nu_{j}}\bigg|^{\beta}
\nonumber\\
&\qquad\times 
\prod_{i<j}\big|\nu_{i}-\nu_{j}\big|^{\beta}
\exp-\left\{\frac{\beta N}{4}
\sum_{i}\nu_{i}^{2}
\right\}\\
&=
\int dK\, \exp\left\{-\frac{\beta N}{4}\text{Tr}\,K^{2}+
S(K)
\right\}
\end{align}
with the new effective action 
\begin{align}
S(K)&= \big(1-\frac{\beta}{2}\big)
\text{Tr}\log h'_{\lambda}(K)\nonumber\\
 &\quad+\frac{\beta}{2}\text{Tr}_{\otimes}\log\bigg(\frac{h_{\lambda}(K)\otimes 1-1\otimes h_{\lambda}(K)}{K\otimes 1-1\otimes K}
 \bigg).
 \end{align}
The main difference with the complex Hermitian case (see \eqref{SpH1}) is the occurrence of the single trace term that involves the derivative. Note the formal similarity between the single and the double trace terms: the former can be obtained from the latter in the limit of coinciding eigenvalues. Therefore, we can apply the previous techniques with only minor modifications, as we sketch below.

In order to apply the LVE formalism, we have to derive the effective action with respect to $K$.  As in the previous section, we use the holomorphic functional calculus to introduce resolvents, so that the first derivative is
 \begin{align}
\frac{\partial}{\partial K}\text{Tr}\log h'_{\lambda}(K)
=\big[1+\widetilde{\Sigma}\big]^{-1} \frac{\partial\widetilde{\Sigma}}{\partial K} .
\end{align}
As before, higher derivatives with respect to $K$ act either on $\big[1+\widetilde{\Sigma}\big]^{-1}$ or on $\frac{\partial\widetilde{\Sigma}}{\partial K}$. The net result is a product of derivatives of $\widetilde{\Sigma}$ with respect to $K$,  separated by insertions of $\big[1+\widetilde{\Sigma}\big]^{-1}$.

The latter factor is nothing but the inverse of $h'_{\lambda}(K)$. Since  the functions $h_{\lambda}$ and $k_{\lambda}$ are inverses one of the other, 
 \begin{align}
\big[1+\widetilde{\Sigma}\big]^{-1}=
 \int_{\Gamma_{u}}du\,k'_{\lambda}(u)\,\frac{1}{u-H}.
 \end{align}
In a basis in which $K$ and therefore also $H$ are diagonal, it obeys the bound  \eqref{Rbound} with $R_{i}=\sup(1, |\mu_{i}|)$.

The second term is obtained by deriving $1+\widetilde{\Sigma}$ with respect to $K$. It simply corresponds to the one obtained in the previous section, except that tensor products are replaced by ordinary products. Explicitly, it reads (see \eqref{derivsigma} for comparison)
\begin{align}
\frac{\partial\widetilde{\Sigma}}{\partial K}&=\int_{\Gamma_{v}}dv\,g_{\lambda}(v)\,\bigg(
\frac{1}{(v-K)^{2}}\sqcup\frac{1}{v-K}+\frac{1}{v-K}\sqcup\frac{1}{(v-K)^{2}}
\bigg),
\end{align}
where $\sqcup$ stands for an insertion an insertion of the two indices of the derivative, as before. Higher order derivatives create new  resolvents, separated by insertions $\sqcup$. 

As a consequence, we obtain the same expression as in the complex case, with the following changes:

\begin{itemize}
    \item there is a single eigenvalue index $i$, so that the limit $\nu_{j}\rightarrow v_{j}$ has to be taken;
    \item tensor products are replaced by ordinary matrix products;
    \item double trace vertices are multiplied by $\frac{\beta}{2}$ and single trace ones by $1-\frac{\beta}{2}$;
    \item tree edges can be twisted or untwisted, with a weight given by \eqref{twist}.
\end{itemize}
Therefore, all the bounds on the corner and derivative operators remain valid for single trace operators, up to multiplicative factors that do not depend on $\epsilon$. Let us also note that the contribution of any single trace vertex is suppressed by a power of $1/N$ in the bounds since it involves one instead of two eigenvalues.  

Moreover, on a tree all twisted edges can be untwisted, so that we conclude that the proof we detailed in the previous section for the complex Hermitian case remains valid in the more general case of real symmetric or quaternionic Hermitian matrices, albeit with modified constants.

\section{Appendix: Positivity of the Jacobian}
\label{PosJ}
In this section we prove the positivity of the Jacobian for $\lambda > 0$ in case of Hermitian matrices.
\begin{lemma} For all $\lambda > 0$ the transformation $\frac{\partial H}{\partial K} > 0$.
\end{lemma}
\proof
The eigenvalues $s_i$ of $K$ are real and in the corresponding eigen-basis the \emph{Jacobian} \eqref{JH} can be written as
\begin{eqnarray}
  J_H &=& \exp\Big(\sum_{i, j}\log\frac{s_i\sqrt{T_p(-\lambda s_i^{2p-2})} - s_j\sqrt{T_p(-\lambda s_j^{2p-2})}}{s_i - s_j}\Big)\,.
\label{JH2log}
\end{eqnarray}
When eigenvalues $s_i$ and $s_j$ have different signs, the expression under the logarithm is positive due to the positivity of the Fuss-Catalan function for $\lambda > 0$ \cite{Krajewski:2017thd} and it produces the positive contribution (as a multiplier)
to the total Jacobian $J_H$.
When $s_i$ and $s_j$ have the same sign, we decompose the corresponding contributions to the Jacobian as
\begin{eqnarray}
\nonumber
  &&\exp\Big(\sum_{i, j}\log\Big[\frac{s_i^2 T_p(-\lambda s_i^{2p-2}) - s_j^2 T_p(-\lambda s_j^{2p-2})}{s_i^2 - s_j^2}\Big]\\
  &+&
  \log\Big[\frac{s_i +s_j}{s_i\sqrt{T_p(-\lambda s_i^{2p-2})} + s_j\sqrt{T_p(-\lambda s_j^{2p-2})}}\Big]\Big)\,.
\label{JH3log}
\end{eqnarray}
Here the argument of the last logarithm under the exponent is positive again due to $T_p(-\lambda s_{i}^{2p-2}) > 0$ for $\lambda > 0$.
Using the functional equation \eqref{FCEq}, we rewrite the argument of the first logarithm in \eqref{JH3log} as
\begin{eqnarray}
\nonumber
  \Big(1 + \lambda\frac{s_i^{2p} T_p^p(-\lambda s_i^{2p-2}) - s_j^{2p} T_p^p(-\lambda s_j^{2p-2})}{s_i^2 T_p(-\lambda s_i^{2p-2}) - s^2_j T_p(-\lambda s_j^{2p-2}}\Big)^{-1} =\\
  \Big(1 + \lambda \sum_{k = 0}^{p-1} (s_i^2 T_p(-\lambda s_i^{2p-2}))^{k} (s_j^2 T_p(-\lambda s_j^{2p-2}))^{p-1-k} \Big)^{-1} > 0\,.
\label{JH4log}
\end{eqnarray}
The positivity of \eqref{JH4log} follows from $\lambda > 0$ and $s_{i}^2 T_p(-\lambda s_{i}^{2p-2}) > 0$.
Since all multipliers in the Jacobian are positive, we have $J_H > 0$.
\qed

\subsection*{Acknowledgments}
The work of VS was supported by the FWF Austrian funding agency through the Schroedinger fellowship J-3981.

\end{document}